\documentclass[aps,pre,twocolumn,groupedaddress,showpacs,showkeys]{revtex4}
\usepackage{latexsym}
\usepackage{amssymb}
\usepackage{graphicx}
\newcommand{\W}{8cm}

\begin{document}


\title{Dispersion enhancement and damping by buoyancy driven flows in 2D networks of capillaries}



\author{Maria Veronica D'Angelo$^{1,2}$}\email{vdangelo@fi.uba.ar}
\author{Harold Auradou$^1$}\email{auradou@fast.u-psud.fr}
\author{Catherine Allain$^1$}
\author{Marta Rosen$^2$}
\author{Jean-Pierre Hulin$^1$}\email{hulin@fast.u-psud.fr}

\affiliation{$^1$Laboratoire Fluides, Automatique et Syst{\`e}mes
Thermiques, UMR 7608, Universit{\'e}s Pierre et Marie Curie-Paris 6 et Paris-Sud,
B{\^a}timent 502, Campus Paris Sud, 91405 Orsay
Cedex, France\\$^2$Grupo de Medios
Porosos, Facultad de Ingenieria, Paseo Colon 850, 1063, Buenos
Aires, Argentina}



\date{\today}

\begin{abstract}
The influence of a small relative density difference ($\simeq  3. 10^{-4}$) 
on the displacement of two miscible liquids is studied experimentally in
 transparent $2D$ networks of micro channels with a mean width  $a$  held vertically. Maps of the local relative concentration are  obtained by an optical light absorption technique. Both stable displacements  in which the denser fluid enters at the bottom of the cell and displaces the lighter
one and unstable displacements in which the lighter fluid is injected at
the bottom and displaces the denser one are realized.  
Except at the lowest mean flow velocity U, the average $C(x,t)$ of the relative concentration  
satisfies a  convection-dispersion equation.  The relative magnitude of $|U|$ and of the
 velocity $U_g$ of buoyancy driven fluid motions is characterized by the gravity number
 $N_g = U_g/|U|$.  At low gravity numbers $|N_g| < 0.01$ (or equivalently high P\'eclet numbers $Pe = Ua/D_m > 500$), the  dispersivities $l_d$ in the stable and unstable configurations are similar  with $l_d\, \propto Pe^{0.5}$.  At low velocities ($|N_g| > 0.01$), 
  $l_d$ increases like $1/Pe$ in the unstable configuration ($N_g < 0$) while  it becomes constant and close to the length of individual channels in the stable case ($N_g > 0$). 
Iso concentration lines $c(x,y,t) = 0.5$ are globally flat in the stable 
configuration while, in the unstable case, they display  spikes and troughs
 with an rms amplitude $\sigma_f$ parallel to the flow.
For $N_g > - 0.2$  $\sigma_f$ increases initially with the distance and reaches
a constant limit while it keeps increasing for  $N_g < - 0.2$. 
 A model taking into account  buoyancy forces driving the instability and  the transverse 
 exchange of tracer between rising fingers and the surrounding fluid is suggested and its applicability to previous results obtained in $3D$ media is discussed.
\end{abstract}

\maketitle

\section{Introduction}
\label{intro}
Miscible displacements in porous media are encountered 
  in many  environmental, water supply and industrial
problems~\cite{bear72,dullien91,sahimi95}. Specific types of
miscible displacements, like tracer dispersion, are also usable as
diagnostic tools to investigate porous media heterogeneities at the
laboratory~\cite{charlaix88} or  field scales.
The characteristics of these processes, such as the width and the
geometry of the  displacement front, are often influenced by
contrasts  between the properties of the displacing and displaced
fluids such as  their density~\cite{freytes01,oltean04,menand05,flowers2007}. 
In unstable density contrast
configurations (fluid density increasing  with height), gravity
driven instabilities may appear and broaden the displacement front:
as an unwanted result, this may lead to early breakthroughs of the
displacing fluid. An example of such effects is the infiltration of 
 a dense plume of pollutant  into a saturated medium.

The objective of the present paper is to study experimentally at both
the local and global scales miscible displacements of two fluids of slightly different
densities ($\Delta \rho/\rho \simeq 3 \times 10^{-4}$): of particular interest is
 the influence of buoyancy driven flow perturbations
on the structure and  development of the mixing zone.

In porous media, the characteristic velocity (counted positively for
upwards flow) of  buoyancy driven flow components  is:
\begin{equation}\label{eq:Ug}
U_g = - k \frac{\Delta \rho g}{\mu}
\end{equation}
in which $\Delta \rho$ is the density of the lower fluid minus that of the
upper one, $\mu$  their viscosity and $k$ the permeability of the medium.
 The velocity $U_g$ is the estimation, from
Darcy's law, of  the  flow per unit area induced  by
the difference of the hydrostatic pressure gradients in the two
fluids. The  relative magnitude  of $U_g$ and of the
mean flow velocity $U$ is a key element in the present problem; it
will be characterized by the gravity number~\cite{flowers2007}:
\begin{equation}\label{eq:gravpar}
N_g = -\frac{U_g}{|U|} =  \frac{\Delta \rho g k}{\mu |U|}
\end{equation}
With the above definition of $\Delta \rho$, one has $N_g > 0$
in the stable configuration (denser fluid  below the lighter one)
and $N_g < 0$ in the unstable one.

Recent experiments~\cite{freytes01,oltean04,menand05} have shown
 that, even when the parameter  $N_g$ is quite small, the geometry of the
mixing fronts may still be strongly influenced by buoyancy. 
Variable density flow and transport in
porous media has therefore received an increasing attention, in particular
through theoretical and numerical
investigations~\cite{schincariol97,beinhorn05,johannsen06}.

In the present work, optical measurements of miscible displacements
of fluids of slightly different densities are performed  in a transparent
two-dimensional vertical network of channels with random
widths~\cite{zarcone83}. The experiments combine visualizations at the
pore scale (one of the fluids is dyed) and measurements of the global concentration 
profiles parallel to the mean flow  at different mean velocities $U$.
Comparing displacement processes  in stable and unstable density contrast
configurations  allows one to detect  and characterize the development
and influence of buoyancy driven flows.

Here, the
density contrasts are low enough so that the $|N_g|$  remains small; under such
conditions the hydrodynamic dispersion damps the development of
instabilities. Then the mixing process can be considered as dispersive and is
 well described by the macroscopic convection-dispersion equation classically 
 used for passive tracers with:
\begin{equation}\label{eq:condiff}
\frac{\partial C}{\partial t} = \vec{\nabla}.(\vec{U}. C - \bar D
.\vec{\nabla} C),
\end{equation}
where $C$ is the tracer concentration, $\vec{U}$ the flow velocity
and $\bar D$  the dispersion tensor (all values are averaged
over the gap of the cell);  $\bar D$ is assumed to
reduce to the diagonal components $D_{\parallel}$ and $D_{\perp}$
corresponding to directions respectively parallel and perpendicular
to the mean flow. The values of $D_{\parallel}$ and $D_{\perp}$ 
are determined by two main
physical mechanisms: advection by the velocity
field inside the medium and molecular diffusion (characterized by a
molecular diffusion coefficient $D_m$). The relative magnitude of these two effects
is characterized by the P\'eclet number $Pe=Ua/D_m$ ($a$  is here the average
 channel width). 
 
 Both immiscible displacements~\cite{zarcone83} and tracer 
dispersion~\cite{charlaix88,dangelo07} have
already been measured previously in such models. This latter
work~\cite{dangelo07} uses the same experimental technique and
porous model as the present one but deals with  experimental conditions
 in which the development of
buoyancy driven instabilities is negligible. In this case, the dye
can be considered as an ``ideal'' tracer that does not modify the
fluid properties. In contrast, the  present work deals with the influence
of buoyancy effects on  dispersion: the components of  $\bar D$
depend on $\Delta \rho$ and are larger in the unstable
configuratio.
Similar studies might be performed on $3D$ porous samples using
NMR-Imaging, CAT-Scan, acoustical techniques~\cite{wong99,kretz03}
or Positron Emission Projection Imaging~\cite{tenchine05} but at a
higher cost and/or with strong constraints on the fluid pairs to be
used.

In the present displacement experiments, concentration maps obtained
for a vertical flow
 are compared for different flow velocities and fluid
rheologies. At the global scale, an effective  dispersion
coefficient is determined and its dependence on the flow velocity  is
studied in both stable and unstable density contrast configurations.
At the local
scale, the variation of geometrical front features of different sizes
is analyzed as a function of the flow velocity and of time.
The combination of these local and global data provides both a sensitive detection of
the instabilities and  information on the characteristics of the
 displacement process at different length scales.
 The $2D$ models of porous media used here are  characterized by a random spatial
distribution of the local permeability: the development of the front geometry under the
combined action of these permeability variations
and of destabilizing density contrasts is of particular interest.
\section{Description of experiment} \label{experiment}
\subsection{Experimental setup and procedure}\label{setup}
The experimental system and the technique for analyzing the data
have already been described in reference~\cite{dangelo07}. The model
medium is a  vertical transparent two dimensional square network  of
channels of random aperture~\cite{zarcone83}: it has a mesh size
equal to  $d = 1\,\mathrm{mm}$ and contains $140 \times140$ channels
with a mean length $0.67\,\mathrm{mm}$ and a depth
$0.5\,\mathrm{mm}$. The  width of the channels takes $7$ values
between $0.1$ and $0.6\, \mathrm{mm}$ with a log-normal distribution
and a mean value $a= 0.33\, \mathrm{mm}$. The permeability of the
network is $k = 3.10^{-9}\,\mathrm{m}^2$ ({\it i.e.} $3000$ Darcy).

The model is vertical with its open sides horizontal (see Figure~2 in reference~\cite{dangelo07}). 
Its upper side is connected to a syringe pump sucking the fluids upward out of the model from 
a reservoir inside which the lower side is dipped. Initially, the model is saturated by pumping 
the first fluid of density $\rho_1$ out of the lower reservoir into the model. Then, the pump is 
switched off and the lower side of the model is removed from the liquid bath by lowering the 
reservoir (the connection tubes are shut to avoid unwanted fluid exchange between the
 model and the outside during this process). The reservoir is then emptied completely, filled 
 up by the second fluid of density $\rho_2 = \rho_1 + \Delta \rho$  and raised again until the lower side of the model
  is below the liquid surface. The displacement process is initiated by opening the connection 
  tubes and switching on the pump. This procedure provides a perfectly straight initial front 
  between the two fluids at the beginning of the displacement.

In this work, the absolute value $|\Delta \rho|$ of the density
difference between the  two fluids is constant with $|\Delta
\rho| = 0.3\,\mathrm{kg/m}^3$. For each flow rate
and pair of fluids used, both stable ($\Delta \rho >\ 0$) and
unstable ($\Delta \rho <\ 0$) configurations are studied by swapping
fluids $1$ and $2$.
\subsection{Fluid characteristics} \label{fluidchar}
Newtonian water-glycerol mixtures or shear thinning water-polymer solutions are
used in the experiments.
The  mean flow velocity ranges from $0.005$ to
$2.5\,\mathrm{mm.s}^{-1}$. The injected and displaced fluids are identical
but for Water Blue dye added to one of the solutions, allowing one to both measure
the local concentration optically and to introduce a controllable density
difference between the fluids (note that, since the density contrast is purely due
 to the dye, the optical determination of the local dye concentration also measures
the local density which is proportional to this concentration).
The molecular diffusion coefficient
$D_m = 6.5 \,\times 10^{-4}\,\mathrm{mm}^2\mathrm{.s}^{-1}$ of the dye is practically
the same in the polymer solutions as in pure water.

The Newtonian water-glycerol solution used is obtained by mixing
$60\%$ in weight of glycerol in pure water. Its viscosity is equal
to $\mu = 10^{-2}\,\mathrm{Pa.s}$ at $20\,^{o}\mathrm{C}$ so that  the molecular diffusion
coefficient (proportional to $\mu^{-1}$) is  $D_m^*=0.7 \times 10^{-4}\,\mathrm{mm}^2\mathrm{.s}^{-1}$.

The shear thinning solutions have  concentrations $C_p = 500$ and
 $1000\,\mathrm{ppm}$ of high molecular weight Scleroglucan (Sanofi
Bioindustries) in water. The variation of their effective viscosity $\mu$ with
the shear rate $\dot{\gamma}$ (see Figure~1 of ref.~\cite{dangelo07}) are
well adjusted by the  Carreau function:
\begin{equation}\label{powervisc}
\mu = \frac{1}{(1+
(\dot{\gamma}/\dot{\gamma_0})^2)^{\frac{1-n}{2}}} (\mu_0 -
\mu_{\infty}) + \mu_{\infty}.
\end{equation}
The values of these rheological parameters for the
solutions used in the present work are listed in Table~\ref{tab:tab1}.
 The limiting value $\mu_{\infty}$ cannot be measured directly and is taken
equal to the viscosity of the solvent, {\it i.e. water}, namely:   
$\mu_{\infty}\,=\,10^{-3}\,\mathrm{Pa.s}$.
 In Eq.~(\ref{powervisc}), $\dot{\gamma_0}$ corresponds to the transition
  between a ``Newtonian plateau" domain at low shear rates ($\dot{\gamma}<\dot{\gamma_0}$)
   in which  $\mu = \mu_0$ and a domain
in  which $\mu$ decreases with  $\dot{\gamma}$ following the power law 
 $\mu \propto {\dot{\gamma}}^{(n-1)}$ ($\dot{\gamma}>\dot{\gamma_0}$).
\begin{table}[htbp]
\begin{tabular}{lcccc}
$C_p$ & $n$ & $\dot{\gamma_0}$ & ${\mu}_0$ & $U^*$\\
 ppm          &     & $\mathrm{s}^{-1}$       & $\mathrm{mPa.s}$ & $\mathrm{mm.s}^{-1}$\\
\\
$500$ & $0.38 \pm 0.04$ & $0.077 \pm 0.018$ & $410 \pm 33$ & $0.045$\\
$1000$ & $0.26 \pm 0.02$ & $0.026 \pm 0.004 $ & $ 4500 \pm 340$ & $0.013$\\
\end{tabular}
\caption{Rheological parameters of scleroglucan solutions used
in the flow experiments. $C_p$ refers to the polymer concentration in ppm. 
The transition mean velocity $U^*$ is estimated by assuming a velocity gradient 
equal to $\dot{\gamma_0}$ at the walls of a channel of width $a$ with a parabolic 
velocity profile.}
\label{tab:tab1}
\end{table}

The model is illuminated from the back by a fluorescent light panel
and images are acquired by a $12$ bits high stability digital camera
with a $1030 \times 1300$ pixels resolution (pixel size = $0.16\,
\mathrm{mm}$). Typically, $100$ images are recorded for each
experimental run at time intervals  between $2.5$ and
$700\,\mathrm{s}$. The images are then translated into maps of the
relative concentration $C(x, y, t)$ of the two fluids in the model
using a calibration procedure already described in
refs.~\cite{dangelo07,boschan07} and commonly used~\cite{menand05}.

\subsection{Characteristic parameters of buoyancy driven flows} 
\label{charbuoyantvel}
 For the water-glycerol mixture, all parameters are fixed and the modulus  $|U_g|$ 
 of the buoyant flow velocity (see Eq.(\ref{eq:Ug})) is constant (only its sign changes
  when the fluids are swapped):
the gravity number $N_g$ varies then as $U^{-1}$ with the velocity and the influence of
gravity is largest at the lowest  velocities. 

For  polymer solutions, the effective viscosity varies with the shear rate
following Eq.~(\ref{powervisc}). and the value of 
$N_g$ can only be computed at  low flow velocities $U < U^*$
 for which the shear rate $\dot{\gamma}$ is lower than $\dot{\gamma_0}$
 in the whole model and  $\mu = \mu_0 = cst.$ so that $N_g$ varies again as $U^{-1}$.
At  velocities $U > U^*$ such that the shear rate is larger than $\dot{\gamma_0}$ 
 in some parts of  the model, the viscosity $\mu$ varies spatially and $N_g$ 
 cannot be computed simply.  
However, at high shear rates one has: $\mu \propto dot{\gamma}^{n-1}$. 
Since  $N_g \propto 1/(\mu |U|)$,  its effective value will be 
proportional to $\dot{\gamma}^{-n}$:  although its actual value cannot be computed, $N_g$ also decreases  when $U$ increases in this shear thinning  regime.  
Due to  this globally monotonous variation, the first instabilities will still appear at the lowest velocities and, often, in the Newtonian domain where $\mu = \mu_0$ so that $U_g = -\Delta \rho gk/\mu_0$.

The velocity $U_g$ has been estimated  for the fluids used
in the present work, leading to $|U_g| = 10^{-3},\ 2.10^{-5}$ and $2.10^{-6}\,\mathrm{mm.s}^{-1}$
for respectively the glycerol-water mixture and the $500\,\mathrm{ppm}$ 
and $1000\,\mathrm{ppm}$ polymer solutions (assuming $\mu = \mu_0$).
At the lowest experimental flow velocity  ($U=5.10^{-3}\,\mathrm{mm.s}^{-1}$),  the duration of
the complete saturation of the network is $t \simeq 3.10^4\,\mathrm{s}$. During this
 time lapse, the distance  $U_g.t$ characterizing the growth of the gravitational
 instabilities is  $28\,\mathrm{mm}$ for the water glycerol mixture: this is far above the length $d$ of
 the individual channels of the network and a noticeable influence of gravity
 on the structure of the displacement fronts is thus  expected.
For the polymer solutions, this same distance is less than $0.6\,\mathrm{mm}$, {\it i.e.} smaller than $d$;  the dye should behave therefore in this case like an "ideal" tracer.

For water-glycerol solutions, buoyancy effects should
be  sizable when the  distance $U_g~t$ becomes
larger than $d$: the transition should then take place at an imposed flow velocity
 $U_c = U_g.L/d \simeq 0.14\ mm.s^{-1}$ ($L$ is the network length). This velocity correspond
to the following gravity and P\'eclet numbers :
$N_g^c\simeq0.01$ and $Pe^c \simeq 500$. These predictions will be shown
In Section~\ref{sec:sec-disp} to correspond well to the experimental results. 
\section{Experimental results}
\subsection{Qualitative observations of miscible displacements}
\label{sec:qualit}
\begin{figure}[htbp]
\noindent\includegraphics[width=\W]{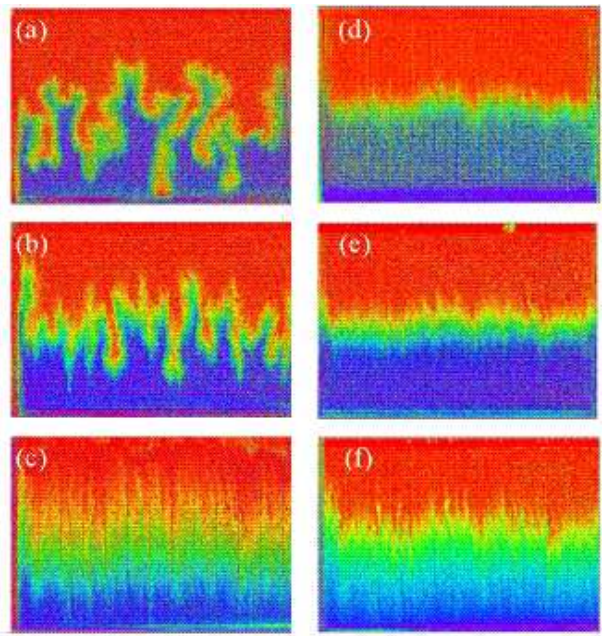}
\caption{Relative concentration maps for experiments using
water-glycerol solutions of different densities at three different
flow rates:  (a, d) $U = 0.005\,\mathrm{mm.s}^{-1}$
 - $|N_g| = 0.2$;
(b, e) $U = 0.025\,\mathrm{mm.s}^{-1}$ - $|N_g| = 0.04$ and (c,f):
$U= 1.25\,\mathrm{mm.s}^{-1}$ - $|N_g| = 8.10^{-4}$. Experiments
correspond to stable (d, e, f) and unstable (a, b, c) density
contrast configurations  at a time when
the injected fluid occupies roughly half of the model. 
In  these figures and in the following ones, darker shades correspond 
to the pure injected or displaced
fluid and the lighter shade to a mixture of the two.
 Fluid flows are upwards with $\vec{g}$ pointing downwards.
 The field of view of the  pictures is $153 \mathrm{mm}\times 140  \mathrm{mm}$.}
\label{fig:fig1}
\end{figure}
Figure~\ref{fig:fig1} displays concentration distributions
observed during displacement experiments using the water-glycerol mixture.
In the stable configuration of Figures~\ref{fig:fig1}~d,e and f, the mean
global front shape remains flat at all flow velocities. The overall
width of the mixing zone  increases however with the flow rate due
to the development of fine structures parallel to the mean flow,
particularly at the highest velocity (Fig.~\ref{fig:fig1}f).

In the unstable configuration,  and at the lowest velocity
(Fig.~\ref{fig:fig1}a), large instability fingers  with a width of
the order of $10$ to $15$ mesh sizes appear and grow up to a length
equal to that of the experimental model.  For a velocity four times
higher, fingers still appear but they are significantly
shorter~(Fig.~\ref{fig:fig1}b).
 As the velocity increases, the size
of the fingers parallel to the mean flow decreases while finer
features develop. For a
velocity still $50$ times higher~(Fig.~\ref{fig:fig1}c), the front
geometry is more similar to that observed in the stable case
although its width parallel to the flow is still broader. In
section~\ref{sec:sec-fronts}, the structure of the mixing zone 
in these experiments will be analyzed again  in the unstable case from
 the geometry of the iso-concentration fronts ($C(x, y, t) = 0.5$).

\begin{figure}[htbp]
\noindent\includegraphics[width=\W]{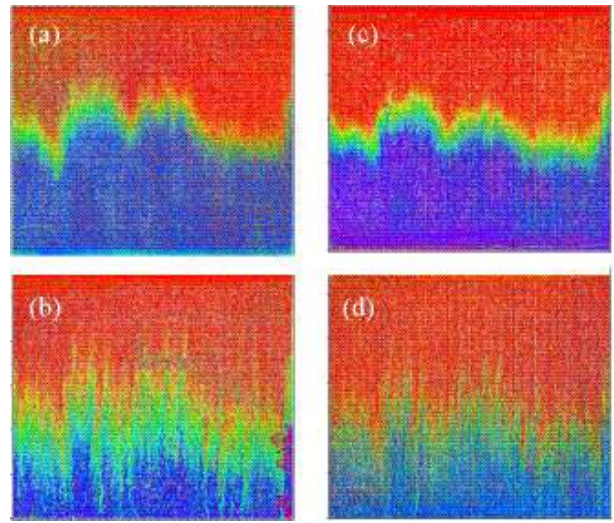}
\caption{Relative concentration maps for experiments using
$1000\,\mathrm{ppm}$ water-scleroglucan solution in gravitationally
unstable (a,b) resp. stable (c,d) configurations for several flow
velocities and gravity number values : $U = 0.005\
mm.s^{-1}$ - $|N_g|= 4.10^{-4}$ (a,c) and $U = 0.5\ mm.s^{-1}$ -
$|N_g|=4.10^{-6}$ (b,d). Grey scale codes and flow direction are the
same as in Figure~\ref{fig:fig1} and the field of view is also identical.}
 \label{fig:fig2}
\end{figure}

Similar pictures from experiments realized  with a $C_p =
1000\,\mathrm{ppm}$ water-scleroglucan solution are displayed in
Figure~\ref{fig:fig2}. In contrast with the case of the
water-glycerol solutions, no buoyancy induced instability appears in
the unstable configurations (cases a,b), even at low velocities (a)
and the relative concentration maps in both configurations are very
similar. At high velocities, fine structures
similar to those observed for the Newtonian solution appear on the
front (cases a,c). Finally, comparisons of front geometries observed
for $C_p = 500$ and $1000\,\mathrm{ppm}$ and displayed in Figure~3
of ref.~\cite{dangelo07} show that intermediate scale structures of
the front (with a width of 10-20 mesh sizes) appear at the same
locations for both concentrations.

Globally, the above results are in qualitative agreement with
macroscopic dispersion
measurements realized on three-dimensional bead packs~\cite{freytes01}
using conductivity tracers detected at the outlet of the samples.  In
these experiments, buoyancy driven instabilities are also observed at
low velocities for Newtonian water-glycerol solutions in a gravitationally
unstable configuration but  do not appear for water-scleroglucan solutions.
Compared to this latter work, the optical  measurements provide additional information
 on the dependence of the front structures of different sizes on the control parameters.
We examine now the variation of the global dispersion
characteristics as a function of the experimental parameters and of
the configuration of the fluids.
\subsection{Quantitative concentration variation analysis}
\label{concvaran}
\begin{figure}[htbp]
\noindent\includegraphics[width=\W]{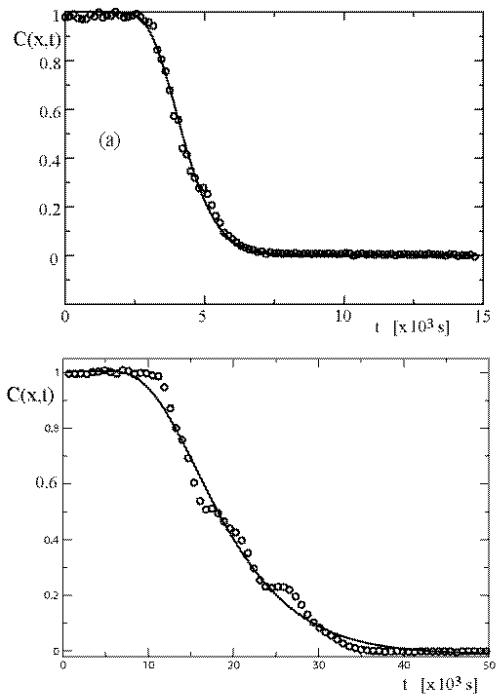} \caption{
Normalized mean concentration variation $C(x,t)$ as a function of
the time $t$  for unstable displacement experiments using a water
glycerol mixture ($C$: average of  local pixel concentration $c(x,
y, t)$ over a width $\Delta y \simeq 144\,\mathrm{mm}$ in the
central part of the model). (a) $U=0.025\,\mathrm{mm.s}^{-1}$,  $N_g
= - 0.04$, $x = 104\,\mathrm{mm}$.
 - (b) $U = 0.005\,\mathrm{mm.s}^{-1}$, $N_g = - 0.2$, $x= 88\,\mathrm{mm}$.
Continuous line: fit by a solution of Equation~(\ref{eq:condiff})
with (a) $\overline{t} = 4350\,\mathrm{s}$, $D_{\parallel}/U^2 =
110\,\mathrm{s}$ and  (b) $\overline{t} = 18266\,\mathrm{s}$,
$D_{\parallel}/U^2 = 1369\,\mathrm{s}$}. The determination of the mean
velocity $U$ is discussed below in the present section~\ref{concvaran}.
\label{fig:fig3}
\end{figure}
\begin{figure}[htbp]
\noindent\includegraphics[width=\W]{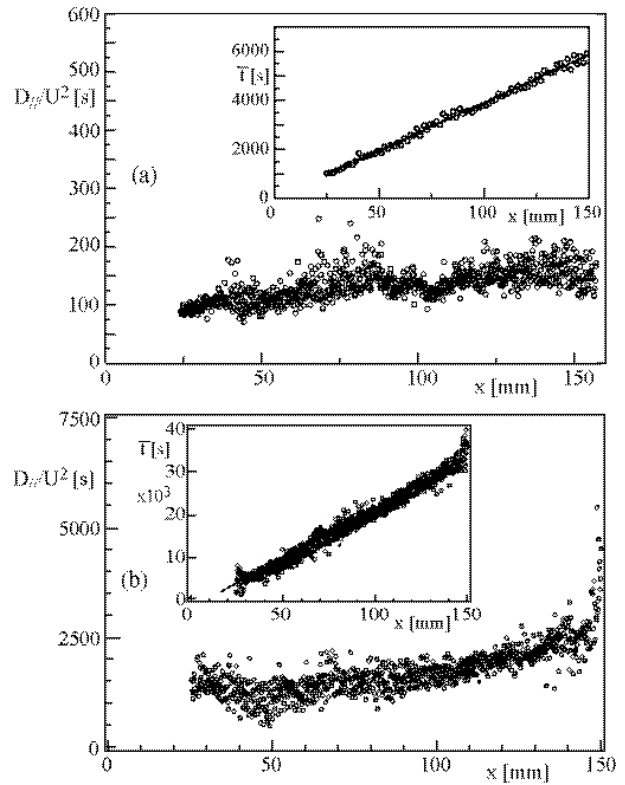}
\caption{Variation of the fitting parameter $D_{\parallel}/U^2$ as a
function of the distance $x$ from the inlet for two unstable
displacement experiments using the water glycerol solutions with $U
= 0.025\,\mathrm{mm.s}^{-1}$ ($N_g = - 0.04$) and $U =
 0.005\,\mathrm{mm.s}^{-1}$ ($N_g = - 0.2$).
 Insets: variation of the mean transit
time $\overline t$ as a function of $x$. Solid line: linear
regression of the data. Values of $\overline t$ and $D_{\parallel}/U^2$
are obtained by fitting the mean concentration variation $C(x,t)$
using Eq.~(\ref{eq:condiff}) at each distance $x$.} \label{fig:fig4}
\end{figure}
The procedure for determining a global dispersion coefficient from
the concentration maps is described in detail in
reference~\cite{dangelo07}. The mean relative concentration   $C(x,t)$ 
of heavy (dyed) fluid at a distance $x$ from the inlet side is
 first determined  by averaging  the value $c(x,y,t)$  for individual pixels over
  a window of width $\Delta y=144\ mm$ representing almost the full width of the model 
  across the flow (only pixels located inside the pore volume are included in the average).
More local information is obtained from the geometry of iso-concentration curves and will be discussed below.

Figures~\ref{fig:fig3}a-b display the variations with time of
$C(x,t)$ for two different values of the gravity number $N_g$.
These variations have been fitted
(continuous line) by  the following one dimensional-solution
of the convection diffusion equation~(\ref{eq:condiff}) (assuming a concentration $C$
constant with y, an initial step-like  variation of $C$ at the inlet and 
taking $D_{\parallel}$ equal to the diagonal component of the tensor $D$ in the $x$ direction) :
\begin{equation}\label{eq:condiffsol}
C(x,t) = \frac{1}{2} \left \lbrack 1 - \frac{N_g}{|N_g|}. \,
erf \left( \frac{t - {\overline t}}{\sqrt{4 D_{\parallel}t/ U^2}}\right)
\right \rbrack
\end{equation}
Since flow is always upwards, adding the factor $N_g/|N_g|  = \pm 1$
makes the equation usable both for light fluid displacing heavy
fluid (unstable configuration - $N_g < 0$) and heavy fluid
displacing light fluid (stable configuration - $N_g > 0$).

For $N_g = - 0.04$, one observes a good fit of the experimental data
with Eq.~(\ref{eq:condiffsol}). For $N_g = - 0.2$ for which the
distortions of the front due to the instabilities are very large,
the concentration does not decrease smoothly but the experimental curve
displays bumps; these features are the signature of rising fingers reaching
the measurement height. Yet, an acceptable fit of the experimental data
with Eq.~(\ref{eq:condiffsol}) can still be achieved. The fits provide the
values of the mean transit time $\overline t$ of the front at the distance $x$ and of
the ratio $D_{\parallel}/U^2 = \bar \Delta t^2/(2\overline t)$
($\bar \Delta t^2$ is the centered second moment of the transit times along the
distance $x$).

\begin{figure}[htbp]
\noindent\includegraphics[width=\W]{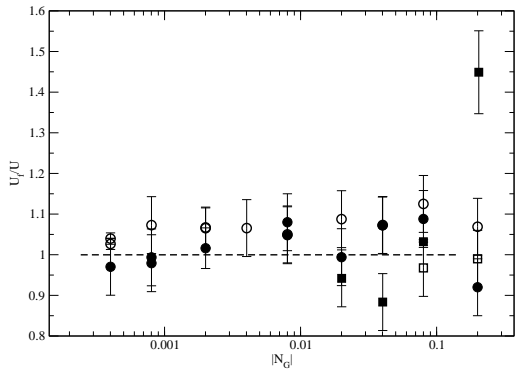} \caption{Variation as a function of $|N_g|$ of the characteristic velocities  of the fluid displacement normalized by the mean velocity $U$ in stable (open symbols) and unstable (dark symbols) density contrast configurations. 
($\circ$),($\bullet$): normalized velocity of the mixing region $U_{mz}/U$   determined  by a linear regression on the variation of $\overline t$ as a function of $x$ (see insets of Fig.~\ref{fig:fig4}a-b). ($\square$, $\blacksquare$) $U_{c0.5}$ = mean velocity of iso-concentration lines $c = 0.5$.} \label{fig:fig5}
\end{figure}

The insets of
figures~\ref{fig:fig5}a-b display the variation of $\overline t$
with the distance $x$ which is linear in both cases; this
 shows that the mixing zone characterized by the mean concentration profile
 $C(x,t)$  moves  at a constant
 velocity $U_{mz}$ which is equal to the inverse of the slope 
 of the variation. A linear regression of
the data provides the values $U_{mz} = 0.027\ \mathrm{mm.s}^{-1}$
 for case $(a)$ and $U_{mz} = 0.0046\ \mathrm{mm.s}^{-1}$ for $(b)$. 
The mean fluid velocity $U$ in the model may be taken equal 
to the ratio of the injected flow rate and of the pore volume per unit length along
$x$ which have been determined independently. 
 The values of $U$ computed in this way at the same two flow rates as above are
  respectively $U=0.025$ and $0.005 \mathrm{mm.s}^{-1}$: they are very close 
to the corresponding values of $U_{mz}$.

The two velocities $U$ and $U_{mz}$ are compared more precisely  in
Figure~\ref{fig:fig6} which displays  the values of 
$U_{mz}/U$ for both stable ($\circ$) and  unstable ($\bullet$) flow configurations,. 
This ratio is always close to $1$: this shows that buoyancy effects do not
influence the mean displacement of the concentration profile, even at the lowest
flow rate ($N_g = -0.2$) for which they are very large (Fig.\ref{fig:fig2}a). One 
assumes therefore in the following that $U = U_{mz}$. 

The main graphics of Figures~\ref{fig:fig4}(a-b) display the variation
 of  the ratio $D_{\parallel}/U^2$ as function of the
  distance $x$ from
the inlet. In case (a) ($N_g = -0.04$), $D_{\parallel}/U^2$ increases
at first slightly  with the distance $x$ and levels off for $x \ge
50\ mm$; then, it  fluctuates  around a constant value.
 Previous studies \cite{dangelo07} have shown that the fluctuations are periodic
  and determined by the structure of the network (the period
  is equal to the mesh size).
 In case (b) ($N_g = -0.2$),
 $D_{\parallel}/U^2$ fluctuates around a constant value for
 $x \le 100\,\mathrm{mm}$.
 At larger distances, a slight steady increase of $D_{\parallel}/U^2$ with $x$ is observed. This is likely due to the rising fingers, directly observable  in Fig.\ref{fig:fig1}a and which can be identified on the curve of Fig.\ref{fig:fig3}b. Tracers are advected faster and farther inside these fingers than outside, resulting in an increase of the dispersivity.
 Even in this case, however, the fluctuations of the measurements and the relatively
 small variations of the values of $D_{\parallel}/U^2$ do not allow one to
 conclude that a diffusive regime is not reached.

These results show both that  the mixing front moves at a constant
velocity $U$ and that (except perhaps
for $N_g = - 0.2$)  it reaches a dispersive spreading regime characterized by a 
dispersion coefficient  $<D_{\parallel}>$  taken equal to the average of 
$D_{\parallel}$ over the full experimental range of $x$ values.
In the following, like in ref.~\cite{dangelo07}, the dispersivity
$l_d\ =\ <D_{\parallel}>/U$ is generally used instead of
$<D_{\parallel}>$ to characterize the dispersion process. Finally,
the standard deviation of  the individual values of $D_{\parallel}$ will be used to estimate the
error bars on $l_d$.
\subsection{Global dispersion measurement results}
\label{sec:sec-disp}
\begin{figure}[htbp]
\noindent\includegraphics[width=\W]{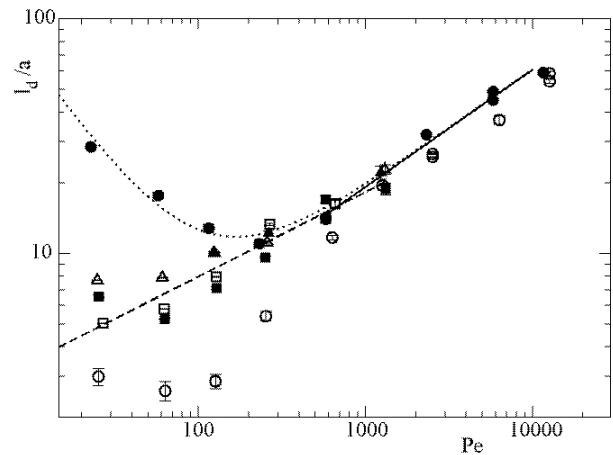}
\caption{Variation of the dispersivity $l_d/a=D/aU$, as a function
of the P\'eclet number $Pe$ for different fluids. $(\circ,\
\bullet)$: water-glycerol mixture; $(\square,\ \blacksquare)$:
$500$\ ppm and $(\triangle, \blacktriangle)$: $1000$\ ppm polymer
solutions. Light symbols: $N_g < 0$; dark symbols: $N_g > 0$. Upper
axis: values of $|N_g|$ for  the Newtonian water-glycerol solutions. Dashed
line: variation law $l_d/a = 1.5 \, Pe^{0.35}$ satisfied from
ref.~\cite{dangelo07} by data corresponding to the polymer solution.
Solid ($l_d/a = 0.6 \, Pe^{0.5}$) and dash-dotted ($l_d/a = 0.5 \, Pe^{0.5}$) lines:, regression for water-glycerol data respectively for the unstable and stable configurations  in the buoyancy free flow domain ($|N_g|<N_g^c = 0.01$). Dotted line: variation predicted by Eq.~(\ref{eq:model}) 
based on the model described in section~\ref{model} below (the dispersivity is assumed
 to be the sum of a passive tracer component and of a buoyancy term which decreasing 
 with $Pe$). The fitting parameters are $\beta_T = 0$ and $\epsilon = 0.3$.}
\label{fig:fig6}
\end{figure}
In the present section, we discuss the  variations of the normalized
dispersivity $l_d/a$
as a function of the flow velocity 
in the stable and unstable configurations: these variations are
displayed in Figure~\ref{fig:fig6}  for the different solutions investigated
(the horizontal scales are the P\'eclet number $Pe = Ua/D_m$ (bottom axis) and the 
gravity number $|N_g|$ (top axis).

For water-glycerol solutions, there is a clear separation at high
$N_g$ (or equivalently low Pe) values between data points  corresponding to unstable ($N_g <
0$) and stable ($N_g > 0$) density contrast configurations: this separation occurs 
occurs close to the transition value $N_g^c = 0.01$ (respectively $Pe^c = 500$) discussed in
Sec.~\ref{charbuoyantvel}. Such a separation has also been previously reported in 3D bead packings~\cite{freytes01}: in Sec.~\ref{model}, these $3D$ observations will be compared quantitatively to the present $2D$ measurements. This separation reflects the
development of finger like structures at low velocities in the
unstable configuration ($N_g < 0$) and  the flattening of the front
in the stable one ($N_g > 0$). These features are due to buoyancy
forces and are qualitatively visible in Figure~\ref{fig:fig6}. At
the lowest P\'eclet number, the values of $l_d$ for $N_g > 0$ and
$N_g < 0$ differ by a factor of nearly $10$.

For $N_g > 0$, the dispersivity $l_d$ reaches for $N_g \simeq 2 \times 10^{-3}$ 
($Pe \simeq 100$) a low minimum value $l_d \simeq 1\,\mathrm{mm}$ close
 to the  characteristic local
length, {\it i.e.} the mesh size of the lattice.
 In most usual homogeneous porous media, and in the
case of a perfectly passive tracer, this latter value represents
 a lower limit of $l_d$: finding such a low value here confirms the
stabilizing influence of the buoyancy forces~\cite{flowers2007}.

For displacements of water-polymer solutions in the same range of
$Pe$ values, the variations of $l_d$ with $Pe$ are the same for the
stable and unstable flow configurations: this result was qualitatively visible
in Figure~\ref{fig:fig2} for the $1000\,\mathrm{ppm}$ solution and
is confirmed quantitatively in Figure~\ref{fig:fig6} for both
polymer concentrations. This implies that no fingering instabilities
develop in the unstable case and that the front is not flattened by gravity
in the stable one. 

Quantitatively, from Sec.~\ref{charbuoyantvel}, an upper limit of the value of $N_g$
can be estimated by taking $\mu = \mu_0$ in Eq.~(\ref{eq:gravpar}) at the
lowest experimental flow velocity. This gives:
 $|N_g| = 5 \times 10^{-3}$ and $|N_g| = 4.5 \times 10^{-4}$ respectively
 for the $500$ and $1000\,\mathrm{ppm}$ solutions, {\it i.e.}
below the threshold of the  instabilities observed for the Newtonian fluids.
Since, $N_g$ decreases at higher flow velocities, this confirms that no effect of buoyancy
will be observable in the experimental range of velocities (in agreement with the experimental observations and with the discussion of Sec.~\ref{charbuoyantvel}). 

At high velocities ($U > 0.1\,\mathrm{mm.s}^{-1}$ (or $Pe > 500$ for water-glycerol and $Pe > 50$ for the polymer solutions),  $l_d$ takes similar values
for both $N_g > 0$ and $< 0$  and increases with $Pe$ as $l_d
\propto Pe^{\alpha}$ with $\alpha = 0.5$ at high $Pe$ values  in both the unstable 
and stable configurations (solid and dash dotted  lines in Figure~\ref{fig:fig6}). 
A similar behaviour has aleady been reported~ref.~\cite{dangelo07} for water 
polymer solutions but with a lower exponent $\alpha = 0.35$: the variation 
 differs in both cases from the
slow increase observed in three-dimensional media such as an homogeneous
monodisperse grain packing~\cite{freytes01}. The $Pe^{\alpha}$
variation reflects the combined effects of geometrical dispersion
due to the disorder of the velocity field and of Taylor dispersion
due to the velocity profiles in individual channels. The latter
becomes important
 at high P\'eclet numbers due to the reduced mixing at the
 junctions by transverse molecular diffusion. As a result, the correlation
 length of the velocity of the tracer particles along their trajectories
  (and, therefore, the dispersivity $l_d$) become larger. The difference
  between the values of $\alpha$ for the Newtonian and shear thinning 
  solutions is likely due to the different relative weight of the two mechanisms in
  the two cases: as discussed in refs.~\cite{dangelo07,boschan07}, Taylor 
  dispersion is indeed reduced and geometrical dispersion is enhanced for shear thinning fluids compared   to the Newtonian case.
 
The above analysis of the global dispersion  discussed  above uses   averages
of the local concentration  over  nearly the full width of the model
and encompassing therefore all the geometrical features of the front
instabilities. The corresponding global value of $l_d$ combines then
different types of effects.

A first one is the local spreading of the displacement front. It is likely to
result from the combined effects the local disorder of the flow field
(geometrical dispersion mechanism) and the flow profile between the rough
walls (Taylor mechanisms).
The effect of these mechanisms in the present system is discussed in ref.~\cite{dangelo07}

A second, more global, effect is the global spreading of the mixing
zone due to the fluid velocity contrasts between different flow
paths (associated, for instance, to the development of the
instabilities). This is studied now
specifically from the variations of the front geometry.
\subsection{Spatial structure of the displacement fronts for unstable
flows} \label{sec:sec-fronts}
Previous experiments in three dimensional porous media~\cite{kretz03}
demonstrated a clear amplification of  front structures resulting
from permeability heterogeneities for unstable density contrast
configurations.
Such effects can be studied  precisely here down
to small lengths scales thanks to the  two-dimensional geometry of the model
network and to  the high precision and spatial resolution of the optical
 concentration measurements.

\begin{figure}[htbp]
\noindent\includegraphics[width=\W]{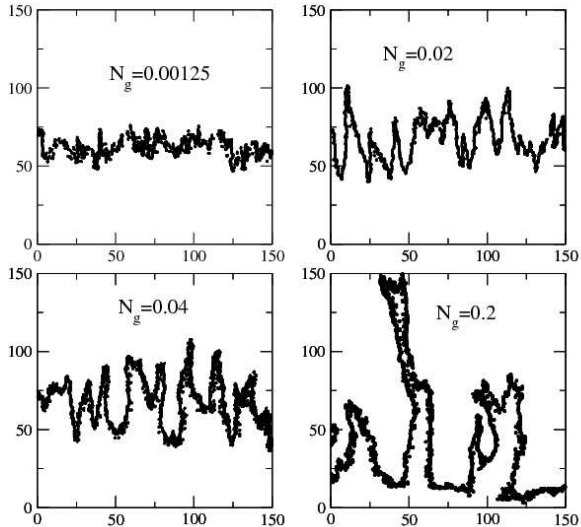} \caption{Iso
concentration fronts $x_f(y,t)$  measured at four different gravity
numbers $N_g = -0.00125$ (a), $- 0.02$ (b), $- 0.04$ (c) and $- 0.2$ (d) for water-glycerol
mixtures in an unstable flow configuration (the injected fluid occupies
half of the model area). Flow is upwards with $\vec{g}$ oriented
downwards. The inset in figure $b$, displays the typical size of the spikes
$\lambda$.} \label{fig:fig7}
\end{figure}
In the following, the front geometry is characterized from the lines
$x_f(y,t)$ along which the local relative concentration $c(x, y, t)$
is equal to $0.5$ at a given time $t$. Examples of such  lines determined by a
thresholding procedure at four gravity numbers $- 0.001 \ge N_g \ge
- 0.2$ are displayed in Fig.~\ref{fig:fig7}.

At the lowest flow velocity investigated ($N_g = - 0.2$, $Pe \simeq 25
$), the buoyancy driven flow components have a major influence on
the front geometry and several instability fingers soar up while the
front displacement is much slower in other regions
(Fig.~\ref{fig:fig7}d).

At higher mean flow velocities (Fig.~\ref{fig:fig7}b-c) the front
retains a rough geometry, still reflecting buoyancy driven flow
components but  its mean advancing motion is clearly visible. 
  In contrast, for $N_g = - 0.2$  (Fig.~\ref{fig:fig7}d),
 the development of the front appears as the combination
of the independent growth of the individual fingers.

 Another important feature is the fact that the distance
$y$ across the front at which corresponding geometrical features
(peaks and troughs) appear remains the same. For $N_g = - 0.2$, the
extension of these features parallel to the flow increases and they
cluster together into larger structures as can be seen in
Figs.~\ref{fig:fig7}b-c. Moreover, for $- 0.02 \ge N_g > - 0.2$, and even though
the width of the front parallel to the mean flow increases significantly, the
typical transverse size $\lambda$ (along $y$) of the individual spikes of the
front is fairly constant wih $\lambda \simeq 4-5 \mathrm{mm}$. This value has
 been taken equal to the mean
 interval (along $y$) between successive local maxima
of the local distance (parallel to $x$) of the front  from the inlet side 
(see insert on figure~\ref{fig:fig7}.b).

These results contrast with the assumptions of a varying
wavelength~\cite{menand05} often applied to porous media following
observations in Hele-Shaw cells~\cite{wooding69}. In the
Hele-Shaw case, there is however no characteristic length scale of the
geometry of the flow field inside the cell, beyond its thickness;
in the present case, on the contrary, the location of the features
of the front appears therefore  be determined by the heterogeneities
of the flow field.

At still higher flow velocities (for instance in
Fig.~\ref{fig:fig7}a) and for $N_g > - N_g^c = - 0.01$ (See
Sec.~\ref{charbuoyantvel} for the expression of $N_g^c$),
distortions of the front due to buoyancy effects decrease in size
and become hardly visible on the iso-concentration lines.

Quantitatively, we characterize these iso-concentration fronts by
their mean position along the flow $\overline{x_f(t)}=<x_f(y,t)>_y$
and by the rms fluctuations $\sigma_f$ of the distance $x_f(y,t)$
($\sigma_f = (<(x_f(y,t)-\overline{x_f(t)})^2>_y)^{1/2}$).

\begin{figure}[htbp]
\noindent\includegraphics[width=\W]{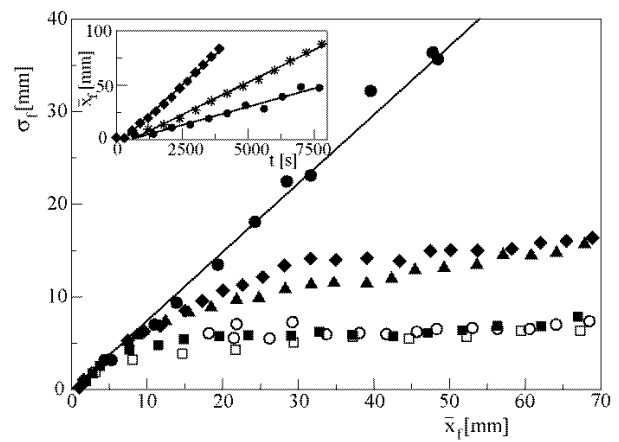}
\caption{Variation of the fronts width $\sigma_f$ as function of the
mean front position $\overline{x_f}$. Empty symbols stand for stable 
configurations, filled symbols for unstable ones.
Squares, triangles, diamonds and circles correspond respectively to
$|N_g|=0.00125$, $0.02$, $0.04$ and $0.2$. Inset: variation of
$\overline{x_f}$ as a function of time in the unstable
configuration for $N_g = -0.2$ ($\bullet$),
$N_g = - 0.08$ ($\ast$) and $N_g = - 0.04$ ($\blacklozenge$). 
Solid lines: linear regressions on data
points.}\label{fig:fig8}
\end{figure}

As shown in the inset of Figure \ref{fig:fig8}, the mean distance
$\overline{x_f(t)}$ increases linearly with time even for the lowest
flow velocity $N_g = -0.2$. The propagation of the iso-concentration 
fronts may therefore be characterized by a  velocity $U_{c0.5}$
determined from a linear regression on the
 variation  of $\overline{x_f(t)}$ with
$t$ ($\square$ and $\blacksquare$ symbols in Fig.~\ref{fig:fig5}).
The velocity $U_{c0.5}$ is very close to the flow  velocity $U$ (see
Sec.~\ref{sec:sec-disp}) for $N_g \geq - 0.08$; however, it  is $40\%$ higher
at the lowest flow velocity ($N_g = - 0.2$). This
confirms other indications of a transition towards a different type of
front growth dynamics (in this latter case, the value of $U_{c0.5}$ is
likely determined by the development of the few large fingers observed in 
Figure~\ref{fig:fig8}).

The variation of $\sigma_f$ with
 $\overline{x_f(t)}$ depends also strongly on the value of $N_g$ (Fig.~\ref{fig:fig8}).
At the highest mean flow velocity ({\it i.e.} $|Ng|= 0.00125$), $\sigma_f$
becomes constant and equal to
$6\,\mathrm{mm}$ as soon as the distance from the injection line is
larger than $10\,\mathrm{mm}$. This value is the same for both
stable and unstable flow configurations: this is in agreement with
the qualitative observations of Sec.~\ref{sec:qualit} in which no
influence of buoyancy is visible when comparing
Figs.~\ref{fig:fig1}c and \ref{fig:fig1}f.

In contrast, at low flow rates (or large $|N_g|$ values), the large
differences between the corresponding concentration maps in
Fig.~\ref{fig:fig1} are reflected in the variations of $\sigma_f$
with $\overline{x_f}$. At the lowest flow-rate ($|N_g| = 0.2$),
$\sigma_f$ keeps increasing roughly linearly with $\overline{x_f}$
in the unstable case; in the stable configuration, in contrast,
$\sigma_f$ reaches quickly a limit of the same order of magnitude as
for the highest velocity. At the
intermediate velocities ($N_g = 0.02-0.04$), the limiting value of
$\sigma_f$ is higher in the unstable configuration than both in the
stable one and at the highest velocity. The distance
$\overline{x_f}$ required to reach this limit is also increased.

This result suggests that, in the unstable configuration and  for intermediate
 $N_g$ values ($- 0.01 > N_g \ge - 0.1$), two distinct regimes are 
 successively observed.
At short distances, the displacement is controlled by the instability while the
variation of $\sigma_f$ is similar to that measured for $N_g = - 0.2$.
At larger distances, the variation of $\sigma_f$ levels off and it reaches a
nearly constant value like for stable displacements. The transition distance
increases with $N_g$: for $N_g = - 0.2$,  it is of the order of the
sample length, explaining why the second regime is not observed.
Together with results  reported previously, we use biw this observation
to provide an analytical estimation of the dispersivity for $N_g < - N_g^c$.

In order to compare the results of the present section~\ref{sec:sec-fronts} with
those of section~\ref{sec:sec-disp}, it must be emphasized that  $\sigma_f$ refers
 to the extension of the iso concentration front $c=0.5$ in the flow
direction: it does not include therefore the influence of the width of the concentration
profile at a given transverse distance $y$. At high velocities, for instance, the iso 
concentration front is nearly flat (Fig.~\ref{fig:fig7}a) and displays only
a few spikes. In this case, $\sigma_f$  reaches a non-zero limit at long distances 
(due to these small features) and it does not increase as $t^{0.5}$ like the
global  width of the mean concentration profile discussed in  Sec.\ref{sec:sec-disp}.
 In the unstable cases and at low velocities, the linear  increase of $\sigma_f$  at long distances reflects directly  the buoyant rise of fingers: its dynamics differs from that of the global spreading of the mixing zone. The latter results  from the combination of several mechanisms (including, but not exclusively, the growth of the fingers)  leading to an increase of the global width of the concenration profile as  $t^{0.5}$. 
\section{Estimation of dispersivity variations for unstable displacements}
\label{model}
The development of fingers driven by buoyancy forces in the unstable configuration
is opposed by lateral mixing induced by transverse dispersion: it
reduces the local
density contrast $\delta \rho$  between the fingers and the surrounding fluid and, finally, the buoyancy
forces. As pointed out in Sec.\ref{sec:sec-fronts}, the structure of
the displacement front has a characteristic size $\lambda$ constant
and close to $4\ mm$ for gravity numbers in the range $- 0.01 > N_ g > - 0.2$.
The characteristic exchange time $\tau$ for lateral mixing
will then be of the order of the transverse diffusion time across the
half-width $\lambda/2$ with:
\begin{equation}
\tau = \frac{\lambda^2}{8 D_\perp}
\label{eq:tau}
\end{equation}
in which $D_\perp$ is the transverse dispersion coefficient.

The rising motion of the fingers driven by buoyancy forces should have a velocity
$u_{finger}$ proportional to the local density contrast with 
$u_{finger} \sim (\delta \rho/\Delta \rho U_g)$. Assuming that $\delta \rho$ decreases
exponentially with a time constant of the order of the transverse mixing time $\tau$
leads to $u_{finger} \sim U_g \, exp(-t/\tau)$.
The vertical displacement
$l(\tau)$ of the rising finger  before its velocity goes to zero
should  then be given by:
\begin{equation}
l(\tau) \simeq  \tau U_g. \label{eq:ltau}
\end{equation}
$l(\tau)$ is then the typical spreading distance of the front due
to buoyancy driven motions during the time  $\tau$: it should
be of the same order of magnitude   as the limiting value of the width $\sigma_f$
of the iso concentration lines  at long distances.
The transverse dispersion coefficient $D_\perp$ generally decreases with the P\'eclet
number (or equivalently with the flow velocity $U$) so that both
 $\tau$ and  $l(\tau)$ should increase at low velocities. This agrees with the observed
  increase of $\sigma_f$ at long distances. At very low velocities, $l(\tau)$ will reache
  (or exceed) the system size: this is indeed observed at the lowest experimental flow
  velocity ($N_g = - 0.2$). In that case, $\sigma_f$ keeps increasing with distance and
does not reach a constant value within the model length.

In order to estimate the dispersion coefficient component
$D_{buoyancy}$ associated to these buoyancy driven motions, one may
consider  $\tau$ represents as the characteristic crossover  time
towards diffusive front spreading: the corresponding width $l(\tau)$
should then verify $l(\tau)^2  \simeq \,2 D_{buoyancy}\tau$.
Combining with Eqs.~(\ref{eq:tau}) and (\ref{eq:ltau}), this leads
to the following value for the dispersivity component $l_{d\
buoyancy}$:
\begin{equation}
  l_{d\ buoyancy} = \frac{D_{buoyancy}}{U} = \frac{U_g^2 \lambda^2}{16 U D_\perp}.
\label{eq:ld_buoyancy}
\end{equation}
As a first approximation, the total
dispersivity $l_d$ is considered to be the sum of the dispersivity  $l_d^{eq}$ for a
fully passive tracer and  of  $l_{d\ buoyancy}$: this amounts to
assume that these two dispersion processes are independent.  This
assumption us only an approximation since the spatial flow velocity
 variations in the model influence both spreading processes.
 
The normalized passive tracer dispersivity $l_d^{eq}/a$ 
 may be estimated from data obtained with the polymer solutions for which, 
as noted above, no buoyancy effect is visible. The equation:
\begin{equation}
\frac{l_d^{eq}}{a} \simeq f Pe^\alpha
 \label{eq:ldeq}
 \end{equation}
where $\alpha=0.35\, \pm \, 0.03$ and $f=1.5\, \pm \, 0.3$ has been selected
for that purpose: it provides indeed a good global fit (dashed line in
Figure~\ref{fig:fig6}) with the polymer data at all $Pe$ values and
an acceptable one with  water-glycerol data either  in the stable configuration
or at high velocities.

Combining Eqs.~(\ref{eq:ld_buoyancy}-\ref{eq:ldeq}) and replacing $U$ by
its expression as a function of $Pe$  leads then to:
\begin{equation}
\frac{l_d}{a}=f Pe^\alpha + \frac{U_g^2 \lambda^2}{16 \epsilon
Pe^{1+\beta_T} D_m^2 } \label{eq:model}
\end{equation}
in which $\epsilon$ is a constant. In this equation  the transverse
dispersion coefficient $D_\perp$ is assumed  to vary as:  $D_\perp =
D_m \epsilon Pe^{\beta_T}$ as suggested by numerical simulations
from ref.~\cite{bruderer01} for networks of capillaries of random radii.

\begin{figure}[htbp]
\noindent\includegraphics[width=\W]{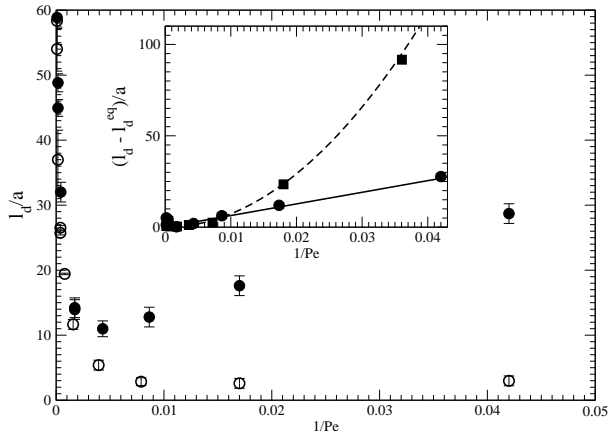}
\caption{Variation of $l_d/a$ as function of $1/Pe$ for the water-glycerol
solution. Filled (resp.
empty) circles correspond to $N_g < 0$ (resp. $N_g > 0$). Inset :
Variation of $(l_d-l_d^{eq})/a$ as function of $1/Pe$ for
unstable experiments in a $2D$ network ($\bullet$) and  a $3D$
porous medium ($\blacksquare$)~\cite{freytes01}. In order to make
comparisons easier, horizontal (resp. vertical) values for $3D$ data
have been divided by a factor $5$ (resp. $0.6$).
Solid (resp. dotted) lines:  regressions using power laws of exponents
$1$ (resp. $2.1$).}\label{fig:fig9}
\end{figure}
In order to put more emphasis on the  buoyancy controlled
regime  at low flow velocities (low $Pe$), the variation of the
dispersivity is plotted in Figure \ref{fig:fig9} as function of
$1/Pe$. For unstable flows, as soon as $1/Pe > 0.0025 \, (1/400)$,
$l_d$ steadily increases with $1/Pe$. This is also the case of the
buoyancy component estimated by subtracting  the passive tracer
dispersivity component $l_d^{eq}$ from the values of $l_d$. The
difference $l_d-l_d^{eq}$ increases linearly with $1/Pe$ and can be
fitted by the second term of Eq.~(\ref{eq:model}) with
$\epsilon=0.3$ and $\beta_T \simeq 0$ (continuous line in the inset
of Figure~\ref{fig:fig9}).

The experimental value of $\beta_T$ is lower than that reported in
ref.~\cite{bruderer01} ($\beta_T = 0.2$); this result corresponds to
numerical simulations for capillary tube networks with a normalized
standard deviation $\sigma_a/a$ of the channel aperture equal to
that of ours
 ($\sigma_a/a = 0.3$). However, in our experiments,  the influence of the
$l_{d\ buoyancy}$ term is mostly significant at the lower P\'eclet
numbers ($Pe \le 50$) while the simulations of
ref.~\cite{bruderer01} deal with $Pe \ge 300$. The variation of
$D_\perp$ for $Pe \le 50$ may therefore be expected to be slower (and the
corresponding exponent lower)  due to the influence of the
 molecular diffusion coefficient $D_m$ which represents a
 constant lower limit at very low P\'eclet numbers.
 
In the present work, the variation of the coefficient of dispersion $D$ results
directly from the 2D nature of the network which has been used.
 In a 3D porous medium, a bead packing for instance,
 $D$ depend weakly on the P\'eclet
number~\cite{dullien91,freytes01,flowers2007}. In this case
$\beta_T$ and $\alpha$ are usually report to be equal and to
range between $1$ and $1.2$~\cite{sahimi95}. As a result, the
difference $l_d-l_d^{eq}$ estimated from the model is expected to
increase with $Pe$ like $\simeq 1/Pe^2$.
We have tested this prediction is tested by re-analyzing the
 dispersion coefficients measured by Freytes et al~\cite{freytes01}.
  In this work, gravity driven instabilities were
studied in a model porous packing of $1\ mm$ glass beads and for a
density difference $\Delta \rho=10^{-3} g/l$. For water and in the
high $Pe$ regime where the effect of  buoyancy is negligible, their data
show that : $l_d^{eq}\propto Pe^{0.1\pm0.1}$ indicating that
$\beta_T \simeq 1.1\pm0.1$. As above, the buoyancy component
is estimated by subtracting the passive tracer dispersivity
$l_d^{eq}$ from the measured dispersivities. The
difference $l_d-l_d^{eq}$ obtained in this way is found to vary as
$(1/Pe)^{1.9\pm0.2}$ (see inset of Fig.~\ref{fig:fig9}) with an
exponent $1.9$ close to the value  $1+\beta_T \simeq 2.1 \, \pm \, 0.1$.
predicted by the model
\section{Conclusions}
To conclude, the miscible vertical displacement measurements on
transparent networks of channels reported here have provided
information at both the local and macroscopic scales on
the mixing front of two miscible fluids of slightly
different densities. Qualitatively, the present macroscopic dispersion
measurement confirm previous ones performed
on three-dimensional porous media~\cite{flowers2007}:
the key feature of this work is that additional new information is provided
by the high resolution visualization of front structures of different sizes
down to the scale of individual channels. Using this information, the front distortions
resulting from instabilities in unstable density contrast configurations could
be analyzed quantitatively.

In these systems, the global spreading of the front
results from the combination of the effects of the disordered spatial variations
of the velocity field (only mechanism active in the
passive tracer case) and of buoyancy driven flows; the latter  may either
decrease or reduce the dispersion depending on the gravitationally stable or
unstable configuration of the fluids. For large P{\'e}clet numbers
($Pe > 500$) ({\it i. e.} small gravity numbers $|N_g| < 0.01$), the displacement
fronts are very similar in both configurations and the global shape of the
iso-concentration fronts is flat. In this case, the front spreading
characteristics and the dispersivity values are similar to those measured
for shear-thinning polymer solutions.

In the stable configuration, the dispersivity decreases significantly
 for $Pe<500$ and becomes constant and close to $1\ mm$ when $Pe<100$.
 At the same time, the geometry of the iso-concentration front is only
weakly affected by the stabilizing effect of buoyancy.

In the unstable configuration, and  at moderate $N_g$ values (e.g.
$- 0.01 > N_g > - 0.2$), the initial development of the instabilities is
damped by transverse hydrodynamic dispersion after some time. As a
result,  front spreading remains dispersive but with  a dispersivity
increasing at low velocities as $l_d \simeq 1/Pe$. These instability fingers are
reflected in the geometry of the iso-concentration fronts which
display large spikes and troughs. In this range of $N_g$ values,
the location (in the transverse direction) and the width of the
spikes are observed to be constant. The spreading in the
flow direction as a function of time displays two regimes: initially (short distances from
the injection), the width of the iso-concentration fronts increases
linearly and, then, it levels-off towards a constant value at large distances.

In order to explain these results, an approach  combining the influences
of longitudinal buoyancy
forces parallel to the mean flow and of the exchange of solute
between the instability fingers and the surrounding fluid has been
developed. It allows one
 to account semi-quantitatively for the dependence of the
dispersivity on the P\'eclet number if $N_g > - 0.2$. The present experimental
observations  on $2D$ networks as well as  previous measurements  on $3D$ bead packings
(for which $l_d\simeq1/Pe^2$) are well fitted by the model.

At the lowest flow velocity investigated ($N_g = - 0.2$), large  fingers
 develop on the interface; the
concentration front is strongly distorted but its spreading can
still be considered as dispersive. Yet, the number of fingers on the
iso-concentration front  decreases with time while its width
parallel to the flow keeps increasing with distance.  This
 suggests that, above this gravity number,  gravitational
instabilities might control the transport process. In 3D porous
media, a linear growth of the mixing zone reflecting the dominant influence
of such
instabilities is only observed above a higher threshold value
$N_g \simeq -1.5$~\cite{menand05}.

Further studies will be needed to confirm our observation
by using pairs of fluids with different density contrasts. Another issue
of practical interest is the influence of the viscosity
contrast on the spatial distribution of the tracer.
\section*{Acknowledgements}
We thank C. Zarcone and the "Institut de M\'ecanique des
Fluides de Toulouse" for realizing and providing us with the
micromodel used in these experiments and G. Chauvin and R. Pidoux for
realizing the experimental setup. This work has been realized in the
 framework of the ECOS Sud program A03-E02 and of a CNRS-CONICET
 Franco-Argentinian "Programme International de Cooperation
 Scientifique" (PICS $n^o 2178$).
%

\end{document}